\documentclass[conference]{IEEEtran}
\usepackage{cite}
\usepackage{amsmath,amssymb,amsfonts}
\usepackage{algorithm}
\usepackage{algorithmic}
\usepackage{graphicx}
\usepackage{textcomp}
\usepackage{xcolor}
\usepackage{bbm}
\usepackage{booktabs}
\usepackage{subfig}
\def\BibTeX{{\rm B\kern-.05em{\sc i\kern-.025em b}\kern-.08em
    T\kern-.1667em\lower.7ex\hbox{E}\kern-.125emX}}

\definecolor{C1}{HTML}{1F77B4}
\definecolor{C2}{HTML}{FF7F0E}

\usepackage{float}
\usepackage{tikz}
\usepackage{pgfplots}
\usepgfplotslibrary{groupplots}
\usepgflibrary{plotmarks} 
\usepgfplotslibrary{colorbrewer} 

\usepgfplotslibrary{external} 

\tikzset{>=latex,
every node/.append style={font=\footnotesize, inner sep=2pt, black, opacity=1},
every pin edge/.append style={<-, very thin, solid, black, opacity=1}
}

\pgfplotsset{
compat=newest,  
/pgf/bar width=12pt,
major grid style={very thin, Greys-E},  
clip=false,         
width=0.8\columnwidth, 
height=1.6in,           
scale only axis,        
scaled ticks=false,     
ylabel style={align=center},    
xlabel style={align=center},    
every axis legend/.append style={column sep=0.05cm, fill=white, legend cell align=left, font = \footnotesize, align=left},     
every tick label/.append style = {font = \footnotesize},         
axis background/.append style={fill=white}, 
every axis plot/.append style={mark options=solid, font = \footnotesize, thick},   
every node near coord/.append style={
    /pgf/number format/fixed zerofill,
    /pgf/number format/precision=3, anchor=west}
}   

\pgfkeys{/pgf/number format/.cd,fixed relative, precision=5, set thousands separator={}} 

\begin{document}

\title{Exploiting topology awareness for routing in LEO satellite constellations}

\author{\IEEEauthorblockN{Jonas W. Rabjerg, Israel Leyva-Mayorga, Beatriz Soret, and Petar Popovski}
\IEEEauthorblockA{Department of Electronic Systems \\
Aalborg University, 9220, Aalborg, Denmark \\
Email: jrabje16@student.aau.dk, ilm@es.aau.dk, bsa@es.aau.dk, petarp@es.aau.dk }}

\maketitle

\begin{abstract}

Low Earth Orbit (LEO) satellite constellations combine great flexibility and global coverage with short propagation delays when compared to satellites deployed in higher orbits. However, the fast movement of the individual satellites makes inter-satellite routing a complex and dynamic problem. In this paper, we investigate the limits of unipath routing in a scenario where ground stations (GSs) communicate with each other through a LEO constellation. For this, we present a lightweight and topology-aware routing metric that favors the selection of paths with high data rate inter-satellite links (ISLs). Furthermore, we analyze the overall routing latency in terms of propagation, transmission, and queueing times and calculate the maximum traffic load that can be supported by the constellation. In our setup, the traffic is injected by a network of GSs with real locations and is routed through adaptive multi-rate inter-satellite links (ISLs). Our results illustrate the benefits of exploiting the network topology, as the proposed metric can support up to $53$\% more traffic when compared to the selected benchmarks, and consistently achieves the shortest queueing times at the satellites and, ultimately, the shortest end-to-end latency.
\end{abstract}


\section{Introduction}

Low Earth Orbit (LEO) satellite constellations are deployed at altitudes from $600$ to $2000$~km above the Earth's surface. These have attracted significant attention recently as they can be used to provide connectivity to remote areas with no terrestrial cellular infrastructure or to offload data traffic in urban hot spots~\cite{Akyildiz2019}.

The satellites in a LEO constellation are usually organized in several orbital planes~\cite{Leyva-Mayorga2020} and can communicate with ground stations (GS) or user equipment on the Earth surface through the ground-to-satellite link (GSL). On the other hand, communication between satellites takes place through the inter-satellite link (ISL). 
The ISLs can be further divided into intra- and inter-plane ISLs. Intra-plane ISLs communicate neighboring satellites in the same orbital plane. Since the inter-satellite distances within the same orbital planes are mainly fixed, intra-plane ISLs are usually stable and ultra-narrow beams, for example, optical wireless links, can be used~\cite{Motzigemba2019}. 

In contrast, satellites in different orbital planes communicate through inter-plane ISLs. Due to the rapid movement of the satellites, inter-plane ISLs are greatly dynamic and may be affected by Doppler shift. Nevertheless, both intra- and inter-plane ISLs are essential for a constellation to serve as a space backbone network without depending on geostationary satellites or a dense network of GSs.

Routing is another essential functionality for constellations serving as a global backbone.
Routing protocols are responsible for: 1) finding appropriate routes for source-destination pairs according to the selected routing metric and 2) defining the forwarding rules. Routing metrics are essential to routing protocols as they determine the cost of each potential hop towards the destination. Classical examples of routing metrics in terrestrial networks include the number of hops (i.e., hop-count), the expected number of (re)transmissions due to packet erasures, euclidean distance, etc. Once the costs have been determined, shortest path algorithms are used to select the path with the lowest total cost before the packet is transmitted from the source; this approach is called unipath routing. 

Routing in LEO satellite constellations has been investigated for many years now~\cite{Ekici2001}. However, with the advent of the New Space era, there is a renovated interest in satellite routing~\cite{Handley2018, Chien2019, Li2017}. The first set of relevant works on LEO routing are from the times of the initial Iridium launches. For example, Ekici \emph{et al.} ~\cite{Ekici2001} proposed a routing algorithm that exploits the geometry of a symmetric Walker star constellation. Intervals in the latitude of the satellites, known as \emph{logical locations} were used to route the packets, forming rings with satellites from different orbital planes. However, this approach is not efficient for constellations with slight asymmetries, as pointed out in our previous work~\cite{Leyva-Mayorga2021}. These slight asymmetries minimize the risk of collisions between satellites and can be found in most commercial dense LEO constellations in the form of slightly different altitudes of deployment for the orbital planes. On the downside, asymmetries in the constellation greatly complicate the routing problem. Hence, recent studies have incorporated the use of Machine Learning, for example, Deep Reinforcement Learning~\cite{Han2020}. 

To the best of our knowledge, the efforts in previous works have oversimplified the constellation geometry and the ISL connectivity, with the exception of papers studying specific commercial constellations~\cite{Handley2018}. In a general approach, we observe that the characteristics of the constellation introduce two distinctive elements to the routing problem. First, the constellation geometry represents a structured dynamic wireless mesh network. Secondly, the propagation time has a great impact on the overall latency. This is in contrast to terrestrial wireless mesh networks, where the propagation time is negligible when compared to the transmission time (i.e., time to transmit a given number of bits at the selected data rate). This aspect requires special attention in the design of routing protocols for LEO constellations.
Lastly, the location of the ground stations and the considered types of traffic greatly impact the traffic load injected to the constellation and the geographic locations where the traffic is injected.

\begin{figure}[t]
    \centering
    \subfloat[]{\includegraphics[width=\columnwidth]{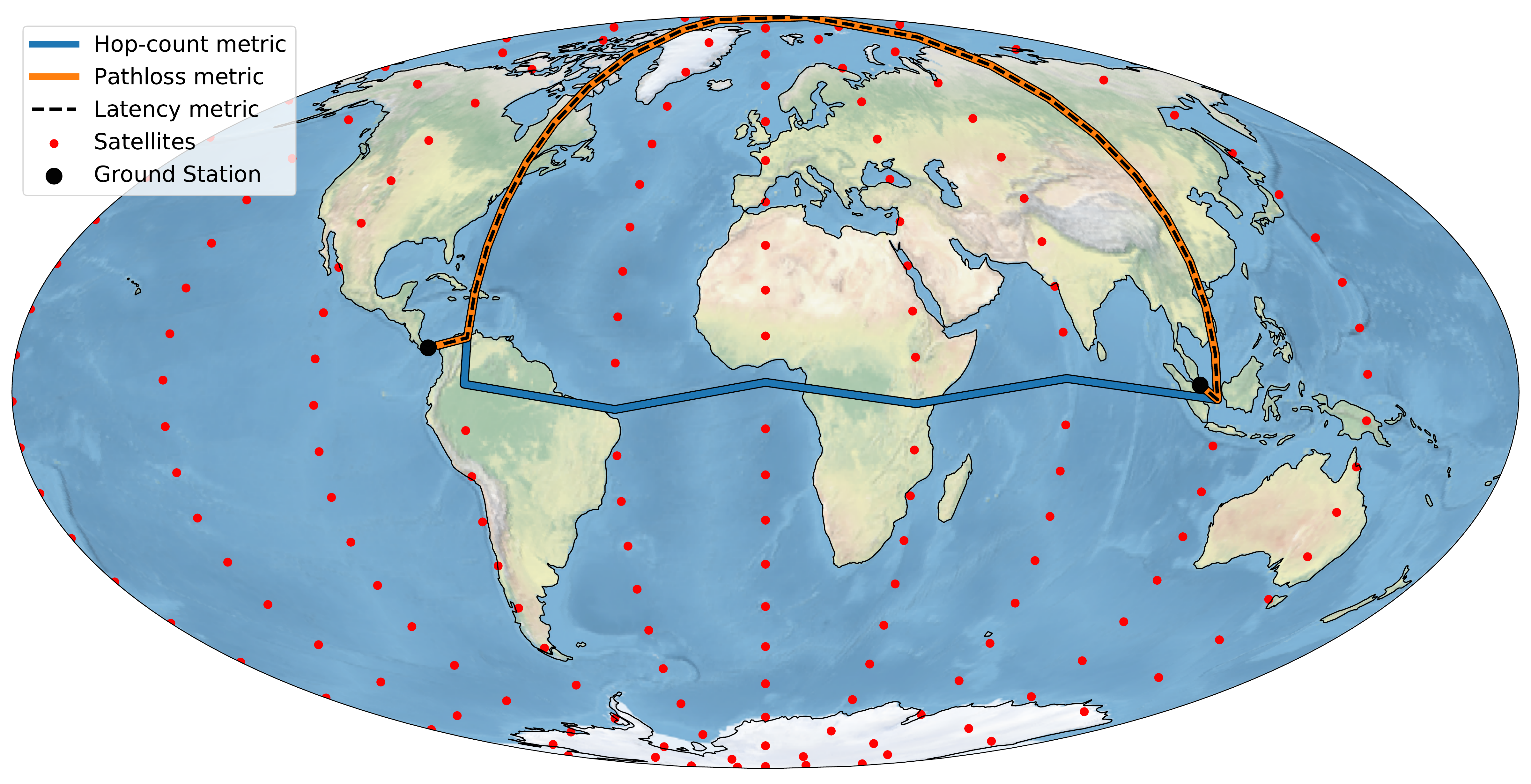}}\\
    \subfloat[]{\includegraphics[width=\columnwidth]{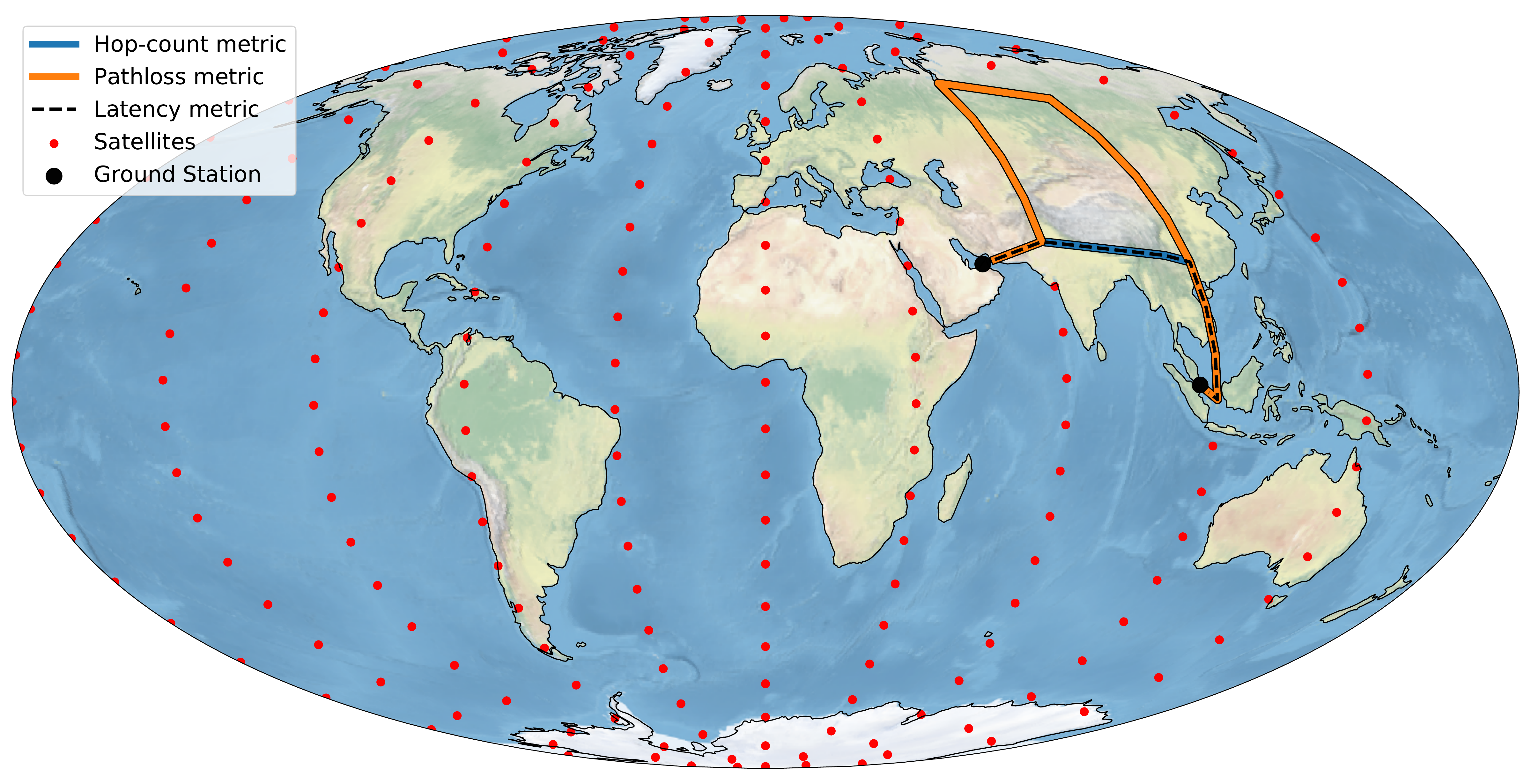}}
    \caption{Diagram of a Walker star constellation with five orbital planes and the routes selected between two distinct GS pairs with three different routing metrics. Made with Natural Earth.}
    \label{fig:constellation_diag}
\end{figure}

In this paper, we consider a scenario where the packets are routed between GSs through a Walker star constellation, creating ground-to-ground logical links~\cite{Leyva-Mayorga2020}. We investigate the performance of unipath routing with multiple rates (i.e., multirate routing). Our analyses focus on 1) the maximum traffic load per GS that can be supported and 2) the routing latency that is achieved in a non-congested Walker star LEO constellation. For this, we propose a new routing metric with low computational complexity that exploits the characteristics of the constellation geometry and favors the selection of hops with high data rates. We simply refer to this metric as the pathloss metric, and its performance is compared to two other relevant routing metrics. First, the hop-count metric, which aims to find the route with a lesser number of ISLs. Second, the latency metric, which aims to find the route with the shortest propagation and transmission times. Fig.~\ref{fig:constellation_diag} illustrates the routes selected with these metrics for two GS pairs. 

Specifically, the main contributions of this paper are
\begin{enumerate}
    \item The description of the pathloss metric, with two variants: normal and low complexity. The low complexity version simply requires knowledge on the polar angles of the satellites and of the general constellation geometry.
    \item A thorough analysis of routing latency in multirate networks, including the impact of diverse traffic flows on the waiting times at each hop.
    \item The analysis of the limitations on the supported traffic load in the constellation with multiple traffic flows and source-destination pairs in a real-life ground infrastructure.
\end{enumerate}

The rest of the paper is organized as follows. Section~\ref{sec:system_model} presents the system model, followed by the calculation of the maximum traffic load per GS that can be supported in Section~\ref{sec:max_arrival} and detailed description of the routing metrics in Section~\ref{sec:routing_metrics}. Then, Section~\ref{sec:results} presents our results and Section~\ref{sec:conclusions} concludes the paper.


\section{System Model} 
\label{sec:system_model}
We consider the routing of packets between ground-station (GS) pairs through a LEO small-satellite constellation~\cite{Leyva-Mayorga2020}. For the ground segment, we consider a set of $N_{\text{GS}}$ ground stations. For the space segment, we consider a Walker star constellation with $M$ polar planes and $N$ satellites. The set of orbital planes is $\mathcal{M}=\{1,2,\dotsc,M\}$; these are evenly separated by $\pi/M$\,radians. The set of satellites is $\mathcal{N}=\{1,2,\dotsc,N\}$. An orbital plane $a\in\mathcal{M}$ consists of $N_a$ evenly spaced satellites deployed at an altitude $h_a$ and with an inclination angle $\epsilon_a=(a-1)\pi/M$. 

While the LEO constellation is dynamic, the time scales for packet transmission are much shorter (in the order of milliseconds) than the orbital periods of the satellites ($>100$ minutes). Hence, we observe the entire system at specific time instants $t\in\mathbb{R}^+$ and skip the time dependence $t$ for notation simplicity. Therefore, the latitude of each satellite $i\in\mathcal{N}$ is simply denoted as $\theta_{i}$ and the coordinates of satellite $i$ in an orbital plane $a$ are $(h_a+r_E, \epsilon_a, \theta_{i})$, where $r_E$ is the radius of the Earth.

Our integrated space and ground infrastructure is modeled as a weighted undirected graph $\mathcal{G}=\left(\mathcal{V}, \mathcal{E}\right)$. Graph $\mathcal{G}$ is multi-partite, with vertex set $\mathcal{V}=\mathcal{U}\bigcup_{a\in\mathcal{M}} \mathcal{V}_a$, where $\mathcal{U}$ is the set of GSs and $\mathcal{V}_a$ is the set of satellites deployed in orbital plane $a$. The edge set $\mathcal{E}$ represents the wireless links established for communication between these. All links are assumed to be half-duplex and used for unicast communication; hence, a transmission queue is maintained for all of them.

The satellites maintain four inter-satellite links (ISLs) whenever possible. Two intra-plane ISLs: one in each direction of the roll axis (aligned with the velocity vector) with the closest intra-plane neighbors, and two inter-plane ISLs: one in each direction of the pitch axis (normal to the orbital plane) with the closest inter-plane neighbors. Therefore, the intra-plane ISLs within an orbital plane $a$ constitute the set of edges $\mathcal{E}_a\subset\left\{ij:i,j\in\mathcal{V}_a^{(2)}\right\}\subset \mathcal{E}$ and $|\mathcal{E}_a|=N_a$. 
On the other hand, the inter-plane ISLs constitute the set of edges $\mathcal{E}_\text{inter}\subset\left\{ij:i\in\mathcal{V}_a, j\in\mathcal{V}_b, a\neq b\right\}\subset \mathcal{E}$. 

Furthermore, the GSs maintain one ground-to-satellite link (GSL) with their closest satellite at all times. These GSLs constitute the set of edges 
\[\mathcal{E}_G \subset \left\{ij:i\in\mathcal{U}, j\in\mathcal{V}_a, a\in\mathcal{M}\right\}.\] 
Finally, we define the edge set as $\mathcal{E}=\mathcal{E}_G\cup\mathcal{E}_\text{inter}\bigcup_{a\in\mathcal{M}}\mathcal{E}_{a}$.
 
The route between GSs $u$ and $v$ is an undirected weighted path denoted as $P_{uv}$. However, the route of a packet from $u$ and $v$ is directed weighted path denoted as $P_{(u,v)}$ and with ordered list of edges $\mathcal{E}\left(P_{(u,v)}\right)=\left(e_1,e_2,\dotsc, e_\ell\right)$, , where $\ell$ is the length of the path. Hence, path $P_{(u,v)}$ is a $u$-path with terminal vertex $v$, where $u$ is an endvertex of $e_1$ and $v$ is an endvertex of $e_\ell$. 

The weights $w(e)$ for all $e\in \mathcal{E}$ are defined by the selected routing metric; the three metrics considered herein are described in the following section. Since the GSLs are the same for each possible path $P_{uv}$ for a given $u$ and $v$, we assume that the GSL links have infinite capacity, which allows us to focus on inter-satellite communication.
  
Inter-satellite communication occurs in a free-space pathloss (FSPL) environment. Let $l(i,j)$ be the slant range (i.e., line-of-sight distance) between two satellites $i$ and $j$. For satellites within line of sight (LoS), $l(i,j)$ is calculated as the euclidean norm between their positions. On the other hand, we set $l(i,j)=\infty$ for the cases with no LoS. Next, let $L_p(i,j)$ be the FSPL and $f$ be the carrier frequency. All the antennas have fixed transmission power $P_t$ and gains in the direction of the main lobe, denoted as $G_t$ for transmission and $G_r$ for reception. Hence, the received signal strength at $j$ from $i$ is
\begin{equation} \label{eq:path_loss}
    P_r(i,j)=\frac{P_t G_t G_r}{L_p(i,j)} =  P_t G_t G_r \left( \dfrac{4 \pi l(i,j) f}{c} \right)^{-2},
\end{equation}
where $c$ is the speed of light and $f$ is the carrier frequency~\cite{Tse}. Throughout this paper, we assume that the satellites have perfect pointing capabilities. Hence, the gain for all the established ISLs is $G_t G_r$. 

It is out of the scope of this paper to design or evaluate interference mitigation techniques and we assume the interference to an ongoing transmission is zero at all times. This can be achieved either by using sufficiently narrow beams, for example, with parameters selected from 3GPP technical reports~\cite{3GPPTR38.821}, or by diverse multiple access techniques that assign orthogonal resources for communication~\cite{Leyva-Mayorga2020}.
Building on this assumption, inter-satellite communication takes place in an interference-free additive white Gaussian noise (AWGN) channel. Hence, the data rate for communication between $i$ and $j$ is selected for a known received power $P_r(i,j)$ from an infinite set of possible values, to be 
\begin{IEEEeqnarray}{rCl}
R(i,j)&=&
B \operatorname{log}_2 \left(1 + \dfrac{P_r (i,j)}{k_B T_s B\gamma} \right),
\label{eq:rates}
\end{IEEEeqnarray}
where $B$ is the bandwidth, $k_B$ is Boltzmann's constant, $T_s$ is the system noise temperature, and $\gamma$ is the signal-to-noise ratio (SNR) margin, selected to avoid outages in the links. 

Knowing the data rate for communication between $i$ and $j$ and the state of the transmission queue at the link $ij$, the exact one-hop latency to transmit a packet of length $p$~bits, simply denoted as $L(i,j)$, can be calculated in the following three parts. 

First, the \textit{waiting time} at the transmission queue $t_w(i,j)$, which is an observation of random variable (RV) $T_w(i,j)$. Second, the \textit{transmission time}, which is the time it takes to transmit $p$~bits at $R(i,j)$~bps. Third, the \textit{propagation time}, which is the time it takes for the electromagnetic radiation to travel the distance $l(i,j)$ from $i$ to $j$. Hence, we have
\begin{equation} \label{eq:latency_metric}
    L(i,j) = \!\underbrace{ t_w(i,j) }_\text{Waiting time}+ \!\underbrace{ \dfrac{ p }{ R(i,j) } }_\text{Transmission time}+\underbrace{ \dfrac{l (i,j)}{c} }_\text{Propagation time}.
\end{equation}

Packets are generated at each GS and transmitted in bursts following a Poisson distribution with rate $\lambda_\text{burst}$. The length of the burst $n$ (in number of packets) and the destination GS are chosen uniformly at random for each burst. Hence, the arrival rate at each GS is equal and denoted as $\lambda=\lambda_\text{burst}\overline{n} p$\,bps, where $\overline{n}$ is the mean length of the burst. A path $P_{(u,v)}$ is selected every time a new burst is generated and remains fixed throughout the transmission of the burst. 

In the following, we provide a recursive expression to calculate the routing latency of a burst with $n$ packets of length $p$ in any path $P_{(u,v)}$ of length $\ell$ with potentially multiple data rates at the edges, denoted as $L(u,v,n,\ell,p)$. 

Let $t_w(e_i,n)$ be the waiting time at the queue of edge $e_i\in\mathcal{E}(P_{(u,v)})$ for the $n$th packet in the burst due to packets from a different traffic flow. 
Next, the routing latency of the first packet through the first $i$ edges of the path $P_{(u,v)}$ is calculated as
\begin{IEEEeqnarray}{c}
L(u,v,1,i,p) = \sum_{m=1}^{i} \left( t_w(e_m,1) +\frac{p}{R(e_m)}+\frac{l(e_m)}{c} \right),\IEEEnonumber\\
\IEEEeqnarraymulticol{1}{r}{\text{s.t. } e_m\in\mathcal{E}(P_{(u,v)})\quad \forall m\leq \ell. }\IEEEeqnarraynumspace
    \label{eq:latency_packet}
\end{IEEEeqnarray}
 The latter formulation corresponds to the routing latency of a single packet, but also defines a set of initial conditions for the recursive calculation. Hence, from the latter, it is easy to calculate the routing latency for the $n$th packet in a burst through the path $P_{(u,v)}$ as
\begin{IEEEeqnarray*}{rCl}
L(u,v,n,\ell,p) &=& \max\Bigg\{L(u,v,n,\ell-1,p),\\
&&\quad L(u,v,n-1,\ell,p)-\frac{l(e_\ell)}{c}\Bigg\}\\
&&+t_w(e_\ell,n)+\frac{p}{R(e_\ell)}+ \frac{l(e_\ell)}{c},\IEEEyesnumber
    \label{eq:routing_latency}
\end{IEEEeqnarray*}
with further initial conditions $L(u,v,n,i,p)=0$ for all $n<0$.
 
\begin{figure}[t]
    \centering
    \includegraphics{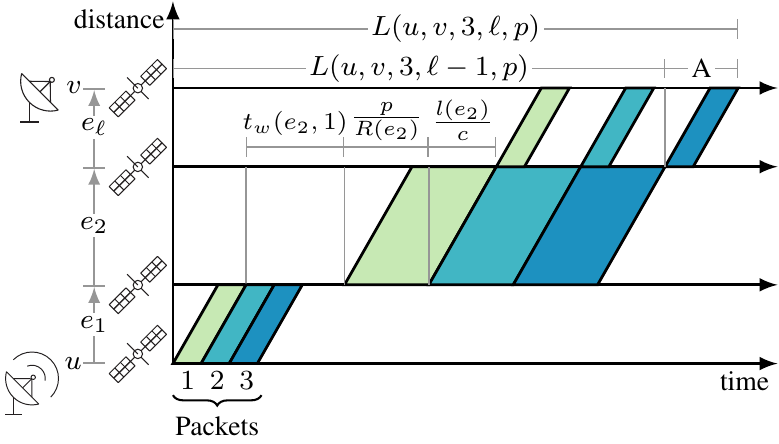}
    \caption{Time diagram for the transmission of three packets along a three-hop satellite route between GSs $u$ and $v$. The satellites are at different distances and use different transmission rates. The length of segment A is $p/R(e_\ell)+l(e_\ell)/c$.}
    \label{fig:time_diagram}
\end{figure}

 The calculation of latency is illustrated in Fig.~\ref{fig:time_diagram}, where a burst of three packets is transmitted along a path with three hops in the satellite constellation. Note that the ISL denoted as $e_2$ has the greatest transmission time and, hence, the greatest contribution to the overall latency. This figure also illustrates that transmitting bursts of $n>1$ packets of $p$\,bits each is more efficient than transmitting all the data in a single packet of size $np$\,bits. The reason for this is that transmission can occur in parallel using different wireless links.

\section{Maximum supported traffic load per GS} \label{sec:max_arrival}
In this section, 
we derive the maximum traffic load per GS that can be supported by the satellite constellation using unipath routing algorithms.
For this, we rely on the Max-flow Min-cut theorem~\cite{Ahlswede2000} to estimate the maximum  traffic load per GS $\lambda_\text{max}(t)$ that can be effectively routed through the satellite constellation at a given time instant. As in the rest of the paper, we remove the time dependency in most of the derivations presented in this section.

 We denote the set of all the selected undirected paths at the same instant as $\mathcal{P}=\left\{P_{uv}\right\}_{\{u,v\}\in\mathcal{U}^{(2)}}$. Since both GSs (i.e., end vertices) in a path $P_{uv}$ generate traffic at rate $\lambda$, the traffic assigned to each path is 
\begin{equation}
    \lambda_P = \frac{2\lambda}{ N_{\text{GS}} - 1}.
\end{equation}

Next, recall that $R(i,j)$ is the data rate selected for communication between satellites $i$ and $j$ selected from the SNR as in\,\eqref{eq:rates} and $R(i,j)=R(j,i)$.
 Therefore, the queues at the ISLs are stable if and only if the following inequality holds
\begin{equation}
    \sum_{P_{uv}\in \mathcal{P}}~\sum_{ij\in \mathcal{E}(P_{uv})} \lambda_P= N_P(ij)\lambda_P\leq R(i,j), \quad \forall i,j\in \mathcal{V},
\end{equation}
where $N_P(ij)$ is the number of paths in $\mathcal{P}$ that contain the edge $ij$. The latter depends on the selected routing metric.

From the Max-flow Min-cut theorem~\cite{Ahlswede2000}, the maximum supported traffic load per GS by the constellation at a specific time instant is
\begin{equation}
    \lambda_\text{max}=\min_{ij\in\mathcal{E}}\frac{R(i,j)\left(N_{\text{GS}} - 1\right)}{N_P(ij)}. \label{eq:lambdamax}
\end{equation}


\section{Routing Metrics} \label{sec:routing_metrics}

In this section we describe in detail the three considered routing metrics. The hop-count and latency metrics are used as benchmark for the pathloss metric. The difference between the routes selected by these is illustrated in Fig.~\ref{fig:constellation_diag} on page~\pageref{fig:constellation_diag}. 

\textbf{Hop-count metric:}
This is a simple routing metric where the weight of each link $ij\in \mathcal{E}$ is $w(i,j)=1$. If two or more paths have the same cost, the selection is made at random. 

\textbf{Latency metric:}
The aim of the latency metric is to deliver the packets using the minimal amount of time. For this, the weight of all edges is set to $w(i,j)=\frac{p}{R(i,j)}+\frac{l(i,j)}{c}$ so that both the propagation and transmission times are considered. On the other hand, the waiting time at the queues is set to $t_w(e)=0$, since their characterization is oftentimes infeasible, at least, at the GSs. The main benefit of the latency metric is that it accounts for the linearity of the propagation times and the non-linearity of the pathloss and, hence, of the achievable data rate $R(i,j)$ as defined in~\eqref{eq:rates}.

\textbf{Pathloss metric:}
This is a relatively simple metric that exploits the constellation geometry  and emphasizes the non-linearity of the pathloss in the ISLs. Hence, it can be easily adapted to the specific constellation geometries.

Let the source GS $u$ be connected to a satellite in orbital plane $a$ and the destination GS $v$ be connected to a satellite in orbital plane $d$. As a starting point, we take the intra-plane ISLs as reference and define the function $f(a, d)$, which takes the value of $1$ if $a\neq d$ and of $\infty$ if $a=d$. Next, since the distance between intra-plane neighbors is greatly similar for all orbital planes, we set $w(i,j)=1$ if $\exists a:ij\in\mathcal{E}_a$. That is, the cost of all intra-plane ISLs is set to $1$. 

On the other hand, the cost of each inter-plane ISL is set to be the ratio of inter- to intra-plane pathloss. Let $i\in\mathcal{V}_a$ be a given satellite in $P_{uv}$ with intra-plane neighbor $j'\in\mathcal{V}_a$ and inter-plane neighbor $j\in\mathcal{V}_b$ s.t. $a\neq b$. Hence, we set
\begin{IEEEeqnarray*}{rCl} 
    w(i,j) &=&\dfrac{ L_p (i,j)f(a, d) }{ L_p (i, j') } \\ 
    &=& \left[f(a, d)\Big(\left(h_a + r_E\right) - \left(h_b + r_E\right)\Big)^2 \right.\\ 
    &&\times\left.\left(\sin \theta_{i} \sin \theta_{j} +\cos \frac{\pi}{ M} \cos \theta_{i} \cos \theta_{j}\right) \right]\\ 
        &&\times\left(\left( 4 (h_a + r_E)^2 \sin^2 \frac{\pi}{N}_a \right) \right)^{-1}\hspace{-1.2em},\quad \forall ij\in\mathcal{E}_\text{inter}. \IEEEeqnarraynumspace\IEEEyesnumber
        \label{eq:relative_pathloss}
\end{IEEEeqnarray*}
Note that $j$ is the closest inter-plane neighbor to $i$ in $b$ if and only if $\theta_{j}\in\left[\theta_{i}-2\pi/N_b,\theta_{i}+2\pi/N_b\right]$.

Therefore, \eqref{eq:relative_pathloss} can be closely approximated by assuming that all orbital planes are deployed at the same altitude $h_a$ and that the satellites are aligned, so that $\theta_{i}=\theta_{j}$, as
\begin{equation} 
     w(i,j)\approx \dfrac{f(a, d) \cos^2 \theta_{i} \left( 1 - \cos \left( \dfrac{\pi}{ M} \right) \right) }{\left(1 - \cos \left( \dfrac{2 \pi}{ N_a } \right)\right)}, \quad\forall ij\in\mathcal{E}_\text{inter}.
     \label{eq:relative_pathloss_simp}
 \end{equation}
Note that the latter approximation greatly reduces the computation complexity as it mostly depends on constant parameters of the constellation, with the exception of $\theta_{i}$. Hence, it can be easily implemented in nodes with low processing capabilities, including LEO satellites. Also note that the weights in this metric greatly depend on $M/N_a$. Therefore, inter-plane ISLs are preferred when the packet is close to the poles, where $\cos^2\theta_{i}\approx 0$, but also when $M>N_a$. Throughout our tests, we observed no difference between the paths selected with \eqref{eq:relative_pathloss} and its low complexity approximation \eqref{eq:relative_pathloss_simp}, hence we use the latter to obtain the results presented in the following section.


\section{Results}
\label{sec:results}
 We consider a Walker star constellation with $M=5$ orbital planes at heights $h_a\geq1000$\,km and $N_a=40$ satellites per orbital plane. The ground segment consists of $N_{\text{GS}} = 23$ GSs placed accordingly to the KSAT ground station service\footnote{https://www.ksat.no/services/ground-station-services/}. 

The relevant parameters and settings are listed in Table~\ref{table:param} along with their default settings; these are used unless otherwise stated.
We have adopted the communication parameters from a recent 3GPP technical report~\cite{3GPPTR38.821}. 


\begin{table}[t]
\centering
\caption{Parameter settings} \label{table:param}
\begin{tabular}{@{}lll@{}}
\toprule
Parameter                   &  Symbol         & Value     \\ \midrule
\# of orbital planes              & $M$       & $5$         \\
\# of satellites per orbital plane  & $N_a$     & $40$         \\
Height of plane $a$~[km]           & $h_a$     &  $1000 + 10(a-1)$  \\
\multicolumn{3}{@{}l}{\textbf{Inter-satellite communication}}\\
\quad Carrier frequency~[GHz]           & $f$     & $20$  \\
\quad EIRP density~[dBW/MHz]        &  $\text{EIRP}_d$   & $4$      \\
\quad Antenna gains~[dB] & $G_t,\,G_r\!$ & $38.5$\\
\quad Bandwidth~[MHz]                   & $B$       & $\{100,400\}$ \\
\quad System temperature~[K]         & $T_s$     & $354.81$    \\
\quad SNR margin~[dB] & $\gamma$ & $2$\\
\multicolumn{3}{@{}l}{\textbf{Ground segment}}\\
\quad \# of GSs & $N_\text{GS}$ & $23$\\
\quad Arrival rate (Mbps)& $\lambda$ & $10$\\
\quad Packet size~(Mbits)     & $p$     & $1$      \\
\quad \# of packets per burst & $n$ & $U(0,20)$\\
\bottomrule
\end{tabular}
\end{table}

Results were obtained by a simulator developed in Python 3.7.6. At each simulation instance, the constellation is rotated randomly with uniform distribution by $\Delta t \sim U (10^4, 10^6)$ seconds according to the orbital velocity of the satellites. Packets are generated at each GS at a rate $\lambda$ for a given period $t_\text{sim}$, which is much shorter than the orbital period of the satellites. During this period, the constellation remains static. For each packet, the destination GS is selected uniformly at random and the route is calculated using Dijkstra's shortest path algorithm with each of the three metrics. 

\begin{figure}[t]
\centering
\begin{tikzpicture}
\begin{groupplot}[
group style={group size=2 by 1, horizontal sep=1cm},
height=1in,
width = 0.37\columnwidth,
xmin=0,
xmax=250,
ymajorgrids,
ymin=0,
ymax=1,
xtick distance=50,
legend style={at={(1,1.07)}, anchor=south east, legend columns =3},
xlabel style={xshift=-0.5em},
restrict x to domain=0:250,
]
\nextgroupplot[ylabel={Empirical CDF}, xlabel={Maximum traffic load $\lambda_\text{max}$~(Mbps)\\[0.5em](a)}]
\addplot[C1] table [x index=0, y index=1] {figures/sourcefiles/cdf_max_flow_100MHz.txt};
\addplot[C2] table [x index=0, y index=2] {figures/sourcefiles/cdf_max_flow_100MHz.txt};
\addplot[black] table [x index=0, y index=3] {figures/sourcefiles/cdf_max_flow_100MHz.txt};
\nextgroupplot[, xlabel={Maximum traffic load $\lambda_\text{max}$~(Mbps)\\[0.5em] (b)}]
\addplot[C1] table [x index=0, y index=1] {figures/sourcefiles/cdf_max_flow_400MHz.txt};
\addplot[C2] table [x index=0, y index=2] {figures/sourcefiles/cdf_max_flow_400MHz.txt};
\addplot[black] table [x index=0, y index=3] {figures/sourcefiles/cdf_max_flow_400MHz.txt};
\legend{Hop-count metric~, Pathloss metric~, Latency metric}
\end{groupplot}
 \end{tikzpicture}
\vspace{-1em}
\caption{Empirical CDF of the maximum traffic load per GS $\lambda_\text{max}$ supported with the three considered metrics with (a) \mbox{$B=100$\,MHz} and (b) $B=400$\,MHz.}
\label{fig:lambda_max}
\end{figure}
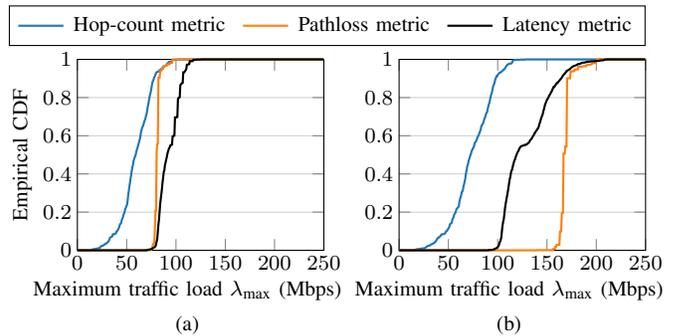

As a starting point, we evaluated $\lambda_\text{max}$, the maximum traffic load per GS that can be served by the constellation, using \eqref{eq:lambdamax}. The empirical cumulative distribution function (CDF) of the values of $\lambda_\text{max}$ obtained with $1000$ distinct rotations of the constellation for the three considered metrics and for $B=\{100,400\}$\,MHz are shown in Fig.~\ref{fig:lambda_max}. As it can be seen, the hop-count metric leads to the lowest values of $\lambda_\text{max}$, which rarely exceed $100$\,Mbps. On the other hand, the pathloss metric supports greater values of $\lambda_\text{max}$ and exhibits a relatively low variance in the results when compared to the other two metrics. Finally, the latency metric achieves mixed results, with a relatively large difference in $\lambda_\text{max}$ between simulation instances. This is clearly observed with $B=400$\,MHz, where in nearly $80$\% of the simulation instances, the $\lambda_\text{max}$ with the latency metric is less than $150$\,Mbps but some instances achieve $\lambda_\text{max}\approx200$\,Mbps.

During our experiments, we observed that the paths calculated with the latency metric are greatly similar to either those with the hop-count metric (with minor variations due to the randomness in path selection) or with the pathloss metric. When the difference in the $\lambda_\text{max}$ between these two metrics is small, as with $B=100$\,MHz, the selection of different paths with the latency metric effectively distributes the traffic load to achieve a greater $\lambda_\text{max}$. On the other hand, when the difference in the $\lambda_\text{max}$ between the hop-count and the pathloss metrics is large, as with $B=400$\,MHz, the $\lambda_\text{max}$ with latency metric oftentimes achieves an intermediate performance.  

 The absolute minimum values obtained for $\lambda_\text{max}$ with $B=400$~MHz are $15.47$\,Mbps with the hop-count metric, $139.57$\,Mbps with the pathloss metric, and $95.61$\,Mbps with the latency metric. Only when $\lambda$ is below these values, the queues of the satellites are guaranteed to be stable. Therefore, routing the packets with our pathloss metric can increase the supported traffic load in the constellation to up to $53$\% with respect to the latency metric and up to $800$\% with respect to the hop-count metric. However, the latency metric achieves a slightly greater arrival rate with $B=100$\,MHz. 

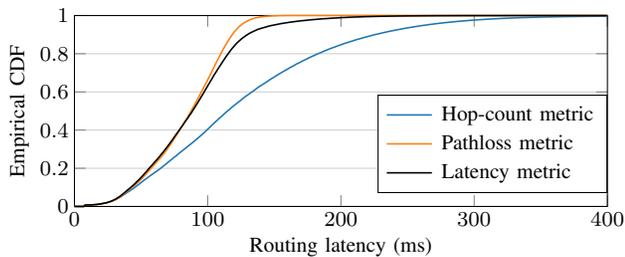
\begin{figure}[t]
\centering
    \tikzsetnextfilename{figs/CDF_latency}
\begin{tikzpicture}
\begin{axis}[
height=1in,
xlabel={Routing latency (ms)},
ylabel={Empirical CDF},
xmin=0,
xmax=400,
ymajorgrids,
ymin=0,
ymax=1,
legend style={at={(1,0.07)}, anchor=south east},
restrict x to domain=0:400
]
\addplot[C1, semithick] table [x index=0, y index=1] {figures/sourcefiles/CDF_latency_1Mbit_400MHz.txt};

\addplot[C2, semithick] table [x index=0, y index=2] {figures/sourcefiles/CDF_latency_1Mbit_400MHz.txt};

\addplot[black, semithick] table [x index=0, y index=3] {figures/sourcefiles/CDF_latency_1Mbit_400MHz.txt};



\legend{Hop-count metric, Pathloss metric, Latency metric}
\end{axis}
 \end{tikzpicture}
\caption{CDF of the routing latency with the hop-count, pathloss, and latency metrics with $B=400$\,MHz.}
\label{fig:routing_latency}
\end{figure}

Next, we show the CDF of the overall packet latency for the three metrics in Fig.~\ref{fig:routing_latency} with $\lambda=10$\,Mbps and $B=400$\,MHz. Here, low percentiles of the CDFs are similar for all metrics and differences between the pathloss metric and the latency metric are only observed for high percentiles. For instance, $0.9$ of the packets are delivered within $120$\,ms with the pathloss metric and within $130$\,ms with the latency metric. Similar conclusions were drawn for the case with $B=100$\,MHz.

\begin{figure}[t]
\raggedright
 \subfloat[]{\tikzsetnextfilename{figures/avg_latency_1mbit}
\begin{tikzpicture}
\begin{axis}[
height=0.8in,
xbar stacked,
area legend,
legend style={ at={(1, 1.05)}, anchor=south east, legend columns=3,/tikz/every even column/.append style={column sep=0.25cm}},
xlabel={Average latency (ms)},
ymin=0.3,
ymax=3.7,
xmin=0,
xmax=200,
xtick distance=50,
xmajorgrids,
minor x tick num=4,
y dir=reverse,
ytick={1,2,3},
yticklabel style={align=center},
yticklabels={{Hop-count}, {Pathloss}, {Latency}}
]
\addplot[fill=Blues-F] table [y=metric,x=propagation] {figures/sourcefiles/Latency_cost_1mbit_100mhz.txt}; 
\addplot[fill=Purples-H] table [y=metric,x=transmission]  {figures/sourcefiles/Latency_cost_1mbit_100mhz.txt}; 
\addplot[fill=Greens-C] table [y=metric,x=queue]  {figures/sourcefiles/Latency_cost_1mbit_100mhz.txt}; 
\legend{Propagation, Transmission, Waiting time}
\end{axis}
 \end{tikzpicture}}\\
 \subfloat[]{\tikzsetnextfilename{figures/avg_latency_1mbit}
\begin{tikzpicture}
\begin{axis}[
height=0.8in,
xbar stacked,
area legend,
legend style={ at={(1, 1.2)}, anchor=south east, legend columns=3,/tikz/every even column/.append style={column sep=0.25cm}},
xlabel={Average latency (ms)},
ymin=0.3,
ymax=3.7,
xmin=0,
xmax=200,
xtick distance=50,
xmajorgrids,
minor x tick num=4,
y dir=reverse,
ytick={1,2,3},
yticklabel style={align=center},
yticklabels={{Hop-count}, {Pathloss}, {Latency}}
]
\addplot[fill=Blues-F] table [y=metric,x=propagation] {figures/sourcefiles/Latency_cost_1mbit_400mhz.txt}; 
\addplot[fill=Purples-H] table [y=metric,x=transmission]  {figures/sourcefiles/Latency_cost_1mbit_400mhz.txt}; 
\addplot[fill=Greens-C] table [y=metric,x=queue]  {figures/sourcefiles/Latency_cost_1mbit_400mhz.txt}; 
\end{axis}
 \end{tikzpicture}}
\caption{Average routing latency per packet due to propagation, transmission, and waiting times for the three metrics with (a) $B=100$\,MHz and (b) $B=400$\,MHz.}
\label{fig:avg_latency}
\end{figure}
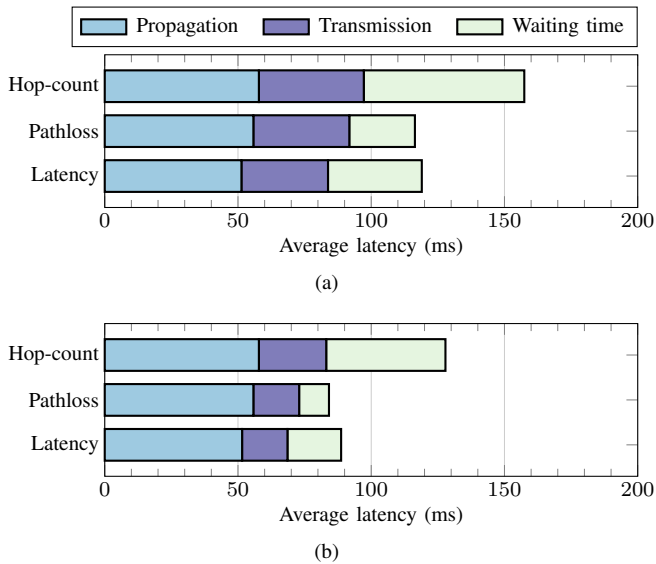

Fig.~\ref{fig:avg_latency} shows the average propagation, transmission, and waiting times per packet with the three metrics with $B=\{100,400\}$\,MHz. The latter reveals the reason why the pathloss metric achieves a faster delivery of the packets in both cases. While the latency metric effectively selects the routes with the shortest propagation and transmission times, the waiting times with the pathloss metric are much shorter. This is because the pathloss metric emphasizes the selection of high data rate links over short routes, which support the greatest traffic load, reduce waiting times, and result in the lowest overall latency. 

\begin{figure}[t]
\centering
    \tikzsetnextfilename{figs/CDF_waiting_time}
\begin{tikzpicture}
\begin{axis}[
height=1in,
xlabel={Waiting time (ms)},
ylabel={Empirical CDF},
xmin=-2,
xmax=50,
ymajorgrids,
ymin=0,
ymax=1,
legend style={at={(1,0.07)}, anchor=south east},
restrict x to domain=0:50
]
\addplot[C1, semithick] table [x index=0, y index=1] {figures/sourcefiles/CDF_waiting_time_1Mbit_400MHz.txt};

\addplot[C2, semithick] table [x index=0, y index=2] {figures/sourcefiles/CDF_waiting_time_1Mbit_400MHz.txt};

\addplot[black, semithick] table [x index=0, y index=3] {figures/sourcefiles/CDF_waiting_time_1Mbit_400MHz.txt};



\legend{Hop-count metric, Pathloss metric, Latency metric}
\end{axis}
 \end{tikzpicture}
\caption{CDF of the waiting time at the queues with the hop-count, pathloss, and latency metrics.}
\label{fig:waiting_time}
\end{figure}
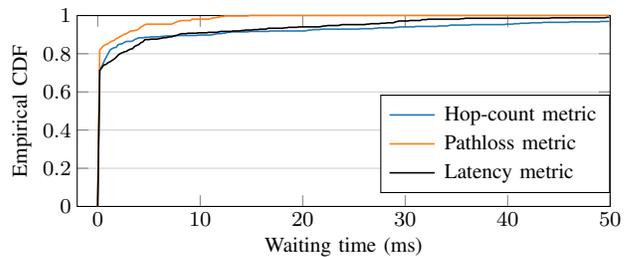

Finally, we show the CDF of the waiting time at the queues for the three metrics with $B=400$\,MHz in Fig.~\ref{fig:waiting_time}. It is important to observe that, since $\lambda<\lambda_\text{max}$, the waiting times at the queues are negligible for most of the packets. Specifically, around $70$\% of the waiting times are $\approx 0$ with all three metrics. However, Fig.~\ref{fig:waiting_time} also shows that some packets experience long waiting times. Specifically, around $10$\% of the waiting times with the hop-count and latency metrics are greater than $10$\,ms.

\section{Conclusion}
\label{sec:conclusions}
In this paper, we proposed a topology-aware routing metric with low computational complexity and evaluated the efficiency of unipath routing with this and two other metrics in a LEO constellation. We considered the use of different rates at each ISL and considered the transmission of multiple packets successively (i.e., bursts). In addition, we derived expressions for the maximum supported traffic load from the GSs.

Our results show that the latency metric can be used find the optimal path, but only in the absence of other traffic flows. On the other hand, our pathloss metric can support a greater traffic load and consistently achieves the shortest latency for the selected transmission parameters. This is due to its emphasis on selecting ISLs with high data rates, which reduces considerably the waiting time at the queues.

Even though a relatively short latency can be obtained with unipath routing, the robustness of the pathloss metric to parameter selection must be further investigated. Besides, the latency metric can be combined with techniques to estimate the waiting time at the queues so these are considered in the routing decision. Besides reducing the latency of individual packets, doing so may greatly increase the supported traffic load by distributing the packets towards idle ISLs.

\bibliographystyle{IEEEtran}

\end{document}